%% file: ms.tex
\documentclass[iop]{emulateapj}             % makes it look like the ApJ format

\usepackage{graphicx}
\usepackage{epsfig}
\usepackage{amsmath}
\usepackage{subfigure} 
\usepackage[colorlinks,citecolor=blue]{hyperref}
\usepackage{amssymb}
\usepackage{apjfonts}
\usepackage{latexsym}
\usepackage{float}
\usepackage{epstopdf}

\newcommand{\TimeScaleRatio}{t_\mathrm{cool}/t_\mathrm{ff}}
\newcommand{\tcool}{t_\mathrm{cool}}
\newcommand{\tff}{t_\mathrm{ff}}
\newcommand{\kb}{k_\mathrm{B}}

\newcommand{\ud}{\,\mathrm{d}}

\newcommand{\Msun}{\text{M}_\odot}

\newcommand{\deriv}[2]{\frac{\ud #1}{\ud #2}}

\shorttitle{Growth and Evolution of Thermal Instabilities}
\shortauthors{G. Meece et. al.}

\begin{document}

\title{Growth and Evolution of Thermal Instabilities in Idealized Galaxy-Cluster Cores}

% Author info
\author{Gregory R. Meece}\footnote{meecegre@msu.edu}
\affil{Department of Physics and Astronomy, Michigan State University,
East Lansing, MI 48824, USA}

\author{Brian W. O'Shea}
\affil{Department of Physics and Astronomy, Michigan State University,
East Lansing, MI 48824, USA}
\affil{Lyman Briggs College, Michigan State University, East Lansing, MI 48825, USA}

\author{G. Mark Voit}
\affil{Department of Physics and Astronomy, Michigan State University,
East Lansing, MI 48824, USA}

% Abstract
\input{abstract}

% Paper contents
\input{intro}
\input{method}
\input{results}
\input{discussion}
\input{conclusions}

% Acknowledgments
\input{acknowledgments}

% Bibliography
\bibliographystyle{apj}
\bibliography{apj-jour,ms}

\end{document}

%% file: abstract.tex
\begin{abstract}
Heat input roughly balances radiative cooling in the gaseous cores of galaxy
clusters even when the central cooling time is short, implying that cooling
triggers a feedback loop that maintains thermal balance.  Furthermore, cores
with short cooling times tend to have multiphase structure, suggesting that the
intracluster medium (ICM) becomes locally thermally unstable for cooling times
$\lesssim 1$~Gyr.  Both observations and theoretical models have linked the
condensation of cold gas with heating by an active galactic nucleus (AGN)
through a cycle in which cooling gas fuels the AGN and drives energetic
outbursts that reheat the ICM and maintain a state of approximate thermal
balance. In this work, we use 2D and 3D hydrodynamic simulations to study the
onset of condensation in idealized galaxy-cluster cores. In particular, we look
at how the condensation process depends on the ratio of cooling time to freefall
time and on the geometry of the gravitational potential.  We conclude that the
ICM can always evolve to a state in which condensation occurs if given enough
time, but that an initial timescale ratio $\TimeScaleRatio \lesssim 10$ is
needed for thermal instability to grow quickly enough to affect realistic
cluster cores within a timescale that is relevant for cosmological structure
formation. We find that instability leads to convection and that perturbations
continue to grow while the gas convects.  Condensation occurs when the timescale
ratio in the low-entropy tail of the perturbation distribution drops below
$\TimeScaleRatio \lesssim 3$, even if the volume-averaged timescale ratio is
substantially greater.  In our simulations, the geometry of the gravitational
potential does not have a strong effect on thermal stability.  Finally, we find
that if condensation is powering feedback, a conversion efficiency of around
$10^{-3}$ for converting the condensed mass into thermal energy is sufficient to
maintain thermal balance in the ICM.
\end{abstract}

%% file: intro.tex
\section{Introduction}\label{section:introduction}

X-ray observations of galaxy clusters have revealed that the radiative cooling
time of gas in many cluster cores is much shorter than the Hubble time. If
radiative cooling were uncompensated by heating, the gas would radiate away its
thermal energy, causing cooling gas to flow toward the center of the cluster.
This would be a classical cooling flow, in which the accumulating cold gas would
be observable and would lead to star formation rates of $\gtrsim 100 \, \Msun \,
{\rm yr}^{-1}$ (see \citet{1994ARA&A..32..277F} for a review). Instead, X-ray
observations reveal little gas cooling below X-ray emitting temperatures
\citep[e.g.][]{2003ApJ...590..207P, 2006PhR...427....1P} and observed
star-formation rates that are one or two orders of magnitude lower than
predicted by the classic cooling-flow model
\citep{2010ApJ...719.1619O,2011ApJ...734...95M}. Thus, an additional process or
processes must be heating the ICM to maintain approximate thermal equilibrium.
Several mechanisms have been proposed and tested through simulations, including
energy injection from supernovae \citep{2007ApJ...668....1N,
2008ApJ...675.1125B, 2013ApJ...763...38S}, conduction of heat from outside of
the core \citep{2002MNRAS.335L...7V,2003ApJ...582..162Z,2013ApJ...778..152S},
heating through mergers \citep{2006NewA...12...71V, 2007PhR...443....1M,
2010ApJ...717..908Z}, dynamical friction from galaxy cluster motion
\citep{2011MNRAS.414.1493R, 2005ApJ...632..157K}, and feedback from AGN
outbursts \citep[reviewed by][]{2007ARA&A..45..117M}, which is the mechanism we
explore in this work.\par

AGN feedback is attractive because a simple order-of-magnitude estimate shows
that an accreting supermassive black hole (SMBH) can easily provide enough
energy to offset cooling. For example, a $10^9$~$\Msun$ SMBH accreting over the
lifetime of the universe and radiating with a mass-energy conversion efficiency
of around 10\% would release a total of $\sim 10^{62}$ ergs, corresponding to an
average power output of around $10^{44}$ ergs per second---easily enough to
offset radiative cooling if a large fraction of that power is injected into the
ICM (see \citet{2002MNRAS.332..729C} for further discussion).  Theoretical and
observational studies support the conclusion that many cool-core clusters host
AGN with enough power to balance cooling \citep[e.g.,][]{2007ARA&A..45..117M,
2006MNRAS.373..959D, 2004ApJ...607..800B} if a significant fraction of the AGN
energy is transfered to the ICM. Nevertheless, the details of the AGN fueling
process and feedback mode are not fully understood. 

If SMBH accretion is to explain thermal regulation of the core, then the
accretion rate must be linked to the thermal properties of the ICM. As pointed
out by \citet{2007ARA&A..45..117M}, if the time-averaged heating rate exceeds
the cooling rate, the core will heat beyond what is observed, and if it is lower
it will fail to prevent gas from cooling. More importantly, the short cooling
times observed in many cluster cores require the heating mechanism to respond on
short timescales, on the order of tens of millions of years. It is therefore
desirable that heating be coupled to the cooling rate, to ensure that feedback
is able to balance cooling both on short timescales and over the lifetime of the
cluster. Two qualitatively different accretion modes have been described in the
literature and implemented in numerical simulations of AGN feedback. Most
implementations base the black hole accretion rate on the properties of the
ambient hot gas using modifications of the classic \citet{1952MNRAS.112..195B}
analysis of smooth, adiabatic accretion, while others rely on condensation and
infall of cold clouds to fuel the black hole
\citep[e.g.,][]{2005ApJ...632..821P,2012ApJ...746...94G,2012MNRAS.424..190G}.
The analysis of \citet{2014arXiv1409.1598V} strongly suggests that the latter
``cold feedback'' mode is more important, because of a universal floor observed
in the radial cooling-time profiles of galaxy clusters that corresponds to the
predicted threshold for condensation of cold clouds
\citep{2012MNRAS.427.1219S}.\par

``Cold mode'' accretion could be fueled by cold gas condensing out of the ICM
in response to thermal instability. The transition of the ICM from a homogeneous
to a hetrogeneous, multiphase structure has a long history of investigation
using theoretical arguments and simulations. From a theoretical standpoint,
\citet{1965ApJ...142..531F} studied the evolution of small perturbations in
cooling plasmas and described an isobaric condensation mode, in which variations
in temperature and density may be amplified. \citet{1970ApJ...160..659D}
extended this analysis, finding that thermal and convective stability are
tightly coupled, a connection further explored in \citet{1989ApJ...341..611B}.
The problem of thermal instability in the context of cooling flows in clusters
was subsequently considered by numerous authors \citep[e.g.][]{1980MNRAS.191..399C,
1986MNRAS.221..377N,1987ApJ...319..632M,1990ApJ...349..471L} who concluded that
the cooling ICM should indeed be subject to thermal instability. However, further
analysis by \citet{1988ApJ...328..395B} and \citet{1989ApJ...341..611B} using a
Lagrangian framework (in contrast to the Eulerian approach of the earlier works)
indicated that the ICM might be less susceptible to thermal instability than
previously thought, especially without the inclusion of a heating term. These
studies generally take as their starting point an equilibrium or steady-state
configuration of gas that may not accurately capture the behavior of the dynamic
ICM. Further, theoretical studies are often incapable of dealing with spatially
dependent heating terms, such as would be expected from star formation and AGN
feedback.\par

There is growing evidence suggesting that the dominant parameter controlling the
transition to a multiphase state and the amount of cold gas that condenses is
the ratio of gas cooling time, $\tcool$, to freefall time, $\tff$. Both numerical
simulations of thermal instability \citep{2012MNRAS.419.3319M} and observations
of galaxy-cluster cores \citep{2008ApJ...683L.107C,
2008ApJ...687..899R,2014arXiv1409.1598V} support this conclusion. Without
gravity to restore equilibrium, multiphase structure can develop within a few
cooling times, because collisional cooling processes scale with the square of
the gas density, allowing denser regions to cool faster than their surroundings.
If the medium is in overall thermal balance, gas clumps that are denser than
average cool and condense, while underdense regions heat and expand faster than
they can cool. In a gravitational potential, however, buoyancy complicates the
development of thermal instability and can inhibit condensation
\citep{1987ApJ...319..632M,1989ApJ...341..611B}. If the freefall time is
significantly longer than the radiative cooling time, an overdense clump can
sink to a denser layer before it can significantly cool.\par

While theoretical studies provide insight into the general physics behind
condensation in the ICM, they are necessarily limited by model assumptions and
can say little about the fate of instabilities that enter the nonlinear regime.
In recent decades, hydrodynamic simulations such as those of
\citet{1990MNRAS.247..367M}, \citet{2012MNRAS.419.3319M}, and
\citet{2014ApJ...789..153L}  have explored the development of thermal
instability in astrophysical environments. These works demonstrated that
condensation can indeed be expected to occur in environments comparable to the
ICM, at a level exceeding the predictions of \citet{1989ApJ...341..611B}.\par

Condensation has been explored in the idealized simulations of
\citet{2012MNRAS.419.3319M}, which show that the growth of thermal instabilities
is significantly inhibited if $\TimeScaleRatio \gtrsim 1$. However, further
studies by \citet{2012MNRAS.420.3174S} have found that in a spherical geometry,
multiphase gas can still condense whenever $\TimeScaleRatio \lesssim 10$ due to
geometric compression (see \citet{2015MNRAS.446.1895S} for further discussion.)
\citet{2012ApJ...746...94G} also finds that a ratio of around 10 is required for
the formation of cold clumps. Alternately, recent work by
\citet{2014ApJ...789..153L} finds that condensation occurs when
$\TimeScaleRatio$ is between 3 and 10. There, condensation is stimulated by
interactions between the ICM and an AGN jet. The jet entrains cold gas from near
the SMBH, pushing it to less dense regions. The clump's positive radial velocity
prevents it from returning to an equilibrium position, and the gas rapidly
cools. Finally, observations by \citet{2015ApJ...799L...1V} find that the
minimum value of $\TimeScaleRatio$ in clusters with multiphase gas in the form
of H$\alpha$ nebulae generally lies between 5 and 30. \par

In this paper, we use idealized 2D and 3D hydrodynamic simulations to study how
the onset of condensation depends on the ratio of cooling time to freefall time
and why there appears to be a change in cluster core properties around a ratio
of 10. Section \ref{section:method} presents simulations based on
\citet{2012MNRAS.419.3319M} in which we explore a wider range of initial
conditions. Section \ref{section:results} analyzes how thermal instabilities
grow in these simulations and investigate how that growth depends on the initial
conditions. Section \ref{section:discussion} relates this work to previous
theoretical work and discusses the validity of these results in the context of
real galaxy clusters. Section \ref{section:conclusions} concludes by discussing
how future work along these lines may clarify how AGN feedback solves the
cooling flow problem in galaxy clusters.\par

%% file: method.tex
\section{Method}\label{section:method}

In this study, we consider simulations of idealized cluster cores with planar,
cylindrical, and spherical geometries in 2 and 3 dimensions.  The simulations
were carried out using the AMR Hydrodynamics code
\texttt{Enzo}\footnote{http://enzo-project.org/} \citep{2014ApJS..211...19B}.
Unless otherwise noted, 2D runs were conducted on a 300x300 cell grid with no
adaptive mesh, and 3D runs employed a $128^3$ cell root grid with 2 layers of
adaptive mesh, with refinement based on overdensity, density gradient, and cooling
time. We do not include magnetic fields, conduction, or the self gravity of the
gas. The simulations were analyzed using the
\texttt{yt}\footnote{http://yt-project.org/} analysis
toolkit \citep{2011ApJS..192....9T}.  \par

\subsection{Problem Setup}

We set up the gas in our simulations subject to the constraint of hydrostatic
equilibrium (HSE) and an `iso-cooling' initial condition, under which the
$\TimeScaleRatio$ ratio is uniform throughout the volume. Additionally, we run a
number of simulations using an isothermal initial condition instead of the
iso-cooling one. \par

The setup described in this section applies to all geometries, as long as the
definition of the height coordinate $z$ changes accordingly.  In planar
geometries, $z$ is the distance from the midplane, in cylindrical geometries it is the
distance from the axis of symmetry, and in spherical geometries it is the distance from the
origin.  We choose a scale height of $z_{\rm S}=100$ kpc (roughly corresponding
to a large cluster), a box size of $R_{\rm S} = 2 z_S$, a scale temperature of
$T_{\rm S} = 10^8$ K, and a gravitational acceleration scale 
\begin{equation}
   g_{\rm S} = \frac{\kb T_{\rm S}}{\mu m_p z_{\rm S}}
\end{equation}
so that the gravitational potential energy and thermal energy are of similar
magnitude at the scale height $z_{\rm S}$. The cooling time is given by 
\begin{equation} \label{equation:cooling_time}
   \tcool(n, T) \equiv \frac{E}{\lvert \dot{E}\rvert} = \frac{3}{2} \frac{n\, \kb T}{n_\mathrm{e}
   n_\mathrm{H} \Lambda(T) }
\end{equation}
where $E$ is the thermal energy per unit volume and the form of the cooling
function $\Lambda(T)$ is taken from \citet{1987ApJ...320...32S} for gas of
half-solar metallicity.\par

The standard normalization of $\Lambda(T)$ is used for gas with an iso-cooling
initial condition of $\TimeScaleRatio = 1$, and we obtain initial conditions
corresponding to other initial values of $\TimeScaleRatio$ by adjusting the
normalization of $\Lambda$ while keeping the gas density and temperature
profiles fixed.  Two time scales characterizing the initial conditions will be
useful in our analysis of the onset of thermal instability.  One is the cooling
time $t_{\rm cool,S}$ at one scale height ($z = z_{\rm S}$) at the beginning of
the simulation. The other is the freefall time at one scale
height, $t_{\rm ff,S}$, which stays constant throughout the simulation.\par

In general, the freefall time of the gas at height $z$ is 
\begin{equation} \label{equation:freefall_time}
  \tff(z) = \sqrt{ \frac {2z} {g(z)} }
\end{equation}
where the gravitational acceleration defining the potential well is

\begin{equation}
  g(z) = - g_{\rm S} \tanh ( \alpha z / z_{\rm S} )
\end{equation}

for $-R_{\rm S} \leq z \leq R_{\rm S}$ and is directed toward either the
midplane, the symmetry axis, or the origin, depending on the geometry of the
potential.  The relative constancy of $g(z)$ away from the origin is meant to
mimic the inner region of a spherical gravitational potential in which the mass
density is proportional to $1/z$.   For $|z| \ll R_{\rm S}$, the $\tanh$
function ensures continuity of the potential, while the parameter $\alpha$
allows adjustment of its cuspiness.  In cylindrical and spherical geometries,
the magnitude of $g(z)$ for $z > R_{\rm S}$ decreases smoothly decreases
smoothly with increasing radius to avoid strong acceleration near the
boundaries. We restrict our analysis to the region $-R_{\rm S} \leq z \leq
R_{\rm S}$.  The simulations we present here use $\alpha=1.0$, which results in
a relatively smooth potential with a gradual softening near the midplane.\par

Following the work of \citet{2012MNRAS.419.3319M}, we implement a heating rate
that exactly balances the average cooling rate at each height. To do this, we
sum the total amount of cooling in each bin of $z$, divide by the total volume
of the bin, and change the sign to get the volumetric heating rate at height
$z$.  While clearly idealized, this heating prescription ensures that the gas
remains in overall thermal balance, in agreement with the observed thermal
behavior of clusters.  The validity of this prescription is discussed in Section
\ref{section:caveats}.\par

For iso-cooling initial conditions, the initial temperature at $z_{\rm S}$ is
$T_{\rm S}$.  Equations \ref{equation:cooling_time} and
\ref{equation:freefall_time} relate density to temperature via
\begin{equation} \label{equation:isocooling_temp_dens}
   \frac {\tcool} {\tff}  = \frac{3}{2} \frac{n\,  \kb T}{\Lambda(T) n_\mathrm{e}
   n_\mathrm{p}} \sqrt{\frac{g(z)}{2\, z}} \; \; ,
\end{equation}
and the HSE condition for an ideal gas is
\begin{equation} \label{equation:short_b}
   \frac {\kb} {\mu m_{\rm p}} 
   \left[ T(z) \frac {\ud \rho} {\ud z} + \rho(z) \frac {\ud T} {\ud z} \right] 
    =  - \rho(z) \, g(z) \; \; 
\end{equation}
Combining these two expressions gives the temperature derivative for iso-cooling
initial conditions in the form of an ODE:
\begin{equation}
   \deriv{\ln T}{\ln z}  = 
                             \left[ \frac { \mu m_{\rm p} g(z) z } { \kb T(z) } 
                                     + \frac {1} {2} \left( \deriv {\ln g} {\ln z} - 1 \right) \right] 
                             \left( \deriv {\ln \Lambda} {\ln T}  - 2 \right)^{-1}
\end{equation}
We integrate this equation to find $T(z)$ and determine the density from the
iso-cooling condition.  For cylindrical and planar simulations, gas outside of
$R_{\rm S}$ is taken to be isothermal and in HSE, with $T(z) = T(R_{\rm S})$.

\begin{figure}[t]
   \includegraphics[width=0.5\textwidth]{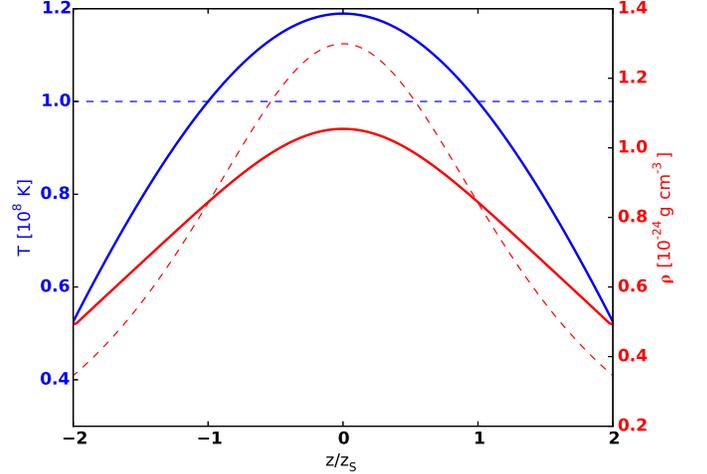}
   \centering
   \caption{The initial temperature (blue) and density (red) profiles used in
   this work is shown for planar geometry. In simulations with cylindrical and
   spherical geometries, the gas is isothermal beyond $z=2$.  The iso-cooling
   setup, with a constant value of $\TimeScaleRatio$, is shown with a solid
   line. The isothermal setup is shown with a dashed line. Both setups are in
   hydrostatic equilibrium and have a ratio of $\TimeScaleRatio=1.0$ at 1 scale
   height. These profiles are used for all simulations; runs with different
   initial values of $\TimeScaleRatio$ are achieved by scaling $\Lambda(T)$
   after initialization.}
   \label{fig:initial_conditions}
\end{figure}

The resulting density and temperature profiles are shown in Figure
\ref{fig:initial_conditions}. 
We impose a temperature floor at $T_{\rm floor} = 5.0 \times 10^6$~K, as we
assume that gas below that temperature inevitably cools rapidly to much lower
temperature.  The details of the gas flow below that temperature occur at finer
resolutions than are employed in our models, and do not affect the overall
condensation rate. Finally, we add randomly generated isobaric perturbations to
the gas with an RMS overdensity of 0.01 and a flat spectrum with wave numbers
between 2 and 20 (with k=1 corresponding to the box size). The same realization of perturbations is used
across all simulations to ensure consistency. As the gas quickly settles into a
convective state, the details of the initial perturbations are soon
forgotten.\par

Figure \ref{fig:initial_conditions} also shows a comparison between the
iso-cooling and isothermal setups. In both setups, the initial density and
temperature profiles do not vary by more than a factor of 4 throughout the
volume. The density is more sharply peaked in the isothermal case, leading to a
shorter cooling time in the center. Consequently, the growth of thermal
instabilities in the isothermal case is more dependent on height and the initial
conditions than in the iso-cooling setup.

%% file: results.tex
\begin{figure*}
   \includegraphics[width=\textwidth]{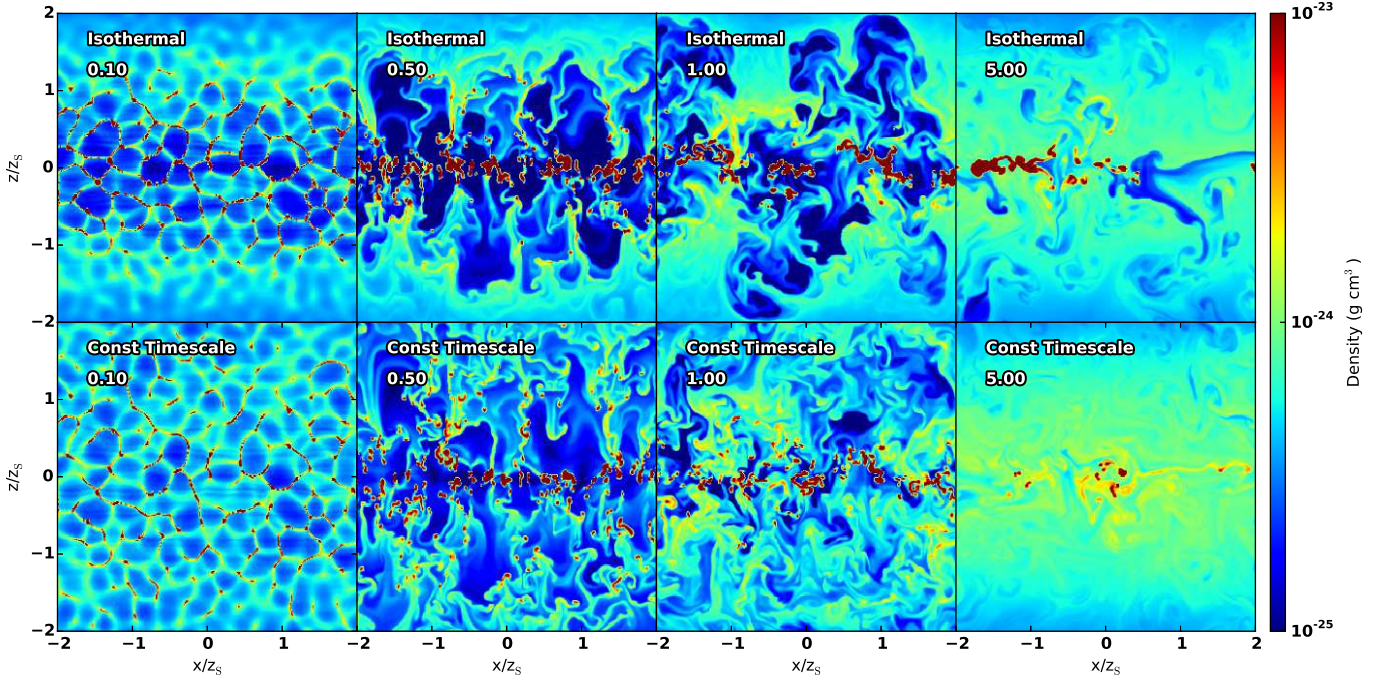}
   \centering
   \caption{ Slices of gas density are shown for 2D planar simulations with initial values
   of $\TimeScaleRatio$ at one scale height of 0.1, 0.5, 1.0, and 5.0 after the
   simulation has evolved for a time $t = 20~t_{\rm cool, S}$. In the top row, the
   gas is initially isothermal. In the bottom row, the initial timescale ratio
   is identical throughout the entire region. Both models produce qualitatively
   similar results. In the isothermal case, gas near the midplane has a shorter
   cooling time than gas above a scale height, leading to earlier condensation
   near the midplane and the creation of hot bubbles that rise up through layers
   that have not yet begun to condense. In both cases, condensation occurs near
   the midplane in simulations with an an initial value of $\TimeScaleRatio = 5$
   at one sale height.}
   \label{fig:mccourt_comparison}
\end{figure*}

\section{Results: The Growth of Thermal Instabilities}\label{section:results}

\subsection{Validation of Method}
We begin by conducting simulations with initial conditions similar to those in
\citet{2012MNRAS.419.3319M} to check if our model produces qualitatively similar
results. Our setup differs from theirs in a number of minor details, including
the shape of the gravitational profile (ours is less cuspy near the midplane)
and the form of the cooling function $\Lambda(T)$.  More importantly, the region
near the midplane does not receive special treatment in our simulations, whereas
\citet{2012MNRAS.419.3319M} shut off heating and cooling in this region and
exclude the midplane region from analysis. In spite of these differences, we
obtain qualitatively similar results in the regime $\TimeScaleRatio \lesssim
1.0$. We see close agreement for the isothermal setup for higher ratios, and
somewhat more condensation is seen for the iso-cooling case for $\TimeScaleRatio
\gtrsim 1.0$. \par

Figure \ref{fig:mccourt_comparison} shows slices of density in 2D Cartesian
simulations with similar initial conditions to those explored by
\citet{2012MNRAS.419.3319M}. The top row of Figure \ref{fig:mccourt_comparison}
uses isothermal initial conditions determined by the initial value of
$\TimeScaleRatio$ at one scale height. In comparison, the simulations in the
bottom row use the iso-cooling initial conditions described in Section
\ref{section:method} for which $\TimeScaleRatio$ is initially constant
throughout the simulation volume. The overall behavior is qualitatively similar
in both cases and resembles the results obtained by \citet{2012MNRAS.419.3319M}
outside of the midplane region.  When the ratio of timescales is below unity,
the gas cools in place and forms droplets of condensate that rain down
towards the midplane. In these cases, convection does not hinder thermal
instability because the gas is able to adjust its thermal state faster than it
is able to convect.  As the ratio of timescales is increased, the dynamics of
the gas become increasingly dominated by convection, although gas continues to
condense around the midplane.\par

While each vertical pair of models illustrated in Figure
\ref{fig:mccourt_comparison} behaves similarly, a number of minor differences
can be observed.  Principally, condensation occurs more uniformly for
iso-cooling initial conditions than for isothermal ones. This result arises from
the differing density profiles needed to satisfy the HSE constraint---isothermal
initial conditions have a steeper gas-density gradient and consequently a larger
range in cooling time across the simulation domain. Shorter cooling times near
the midplane lead to a `cross talk' effect that is more pronounced for
isothermal initial conditions. Condensation of gas near the midplane causes hot,
low-density bubbles to form there and to rise to greater altitudes, creating
inhomogeneity at those altitudes on a freefall time scale instead of a cooling
time scale. This cross talk between lower and upper layers complicates the task
of interpreting how thermal instability and condensation depend on the choice of
initial timescale ratio at one scale height. The `iso-cooling' condition, while not
necessarily more physically valid, reduces this cross talk and allows for
clearer interpretation of the relationship between the initial timescale ratio
and the onset of condensation. In contrast, dense midplane gas in models with
isothermal initial conditions is able to condense quickly even when the initial
ratio of timescales at one scale height is large. This happens because the gas
near the midplane will have a lower ratio of $\TimeScaleRatio$, leading to
localized condensation.\par

\subsection{Instability Growth in the Strong Cooling Regime} 
\begin{figure}
   \includegraphics[width=0.5\textwidth]{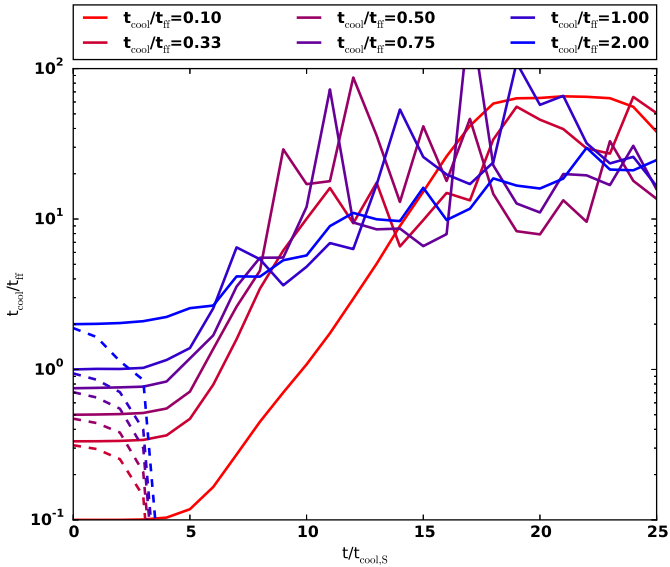}
   \centering
   \caption{The average ratio of cooling time to freefall time in the ambient
   gas is shown as a function of time for simulations with low initial values of
   $\TimeScaleRatio$. The $x$ axis is in units of the initial cooling time at
   one scale height, $t_{\rm cool, S}$, rather than absolute time. Values are
   shown for the 2D planar geometry case. Solid lines indicate the
   volume-averaged value of $\TimeScaleRatio$ within a zone $0.8~z_{\rm S} < z
   <1.2~z_{\rm S}$. Dashed lines show the minimum value of $\TimeScaleRatio$
   within the entire box. At low values of the initial timescale ratio, the gas
   is able to cool in place within a few cooling times, driving the rest of the
   gas to $\TimeScaleRatio > 10$. As the gas cools largely in place,
   instabilities grow purely on the cooling time, leading to similar behavior
   for all runs.} \label{fig:timescale_short}
\end{figure}

Using the iso-cooling simulations presented in the previous section, we have
examined the evolution of perturbations for the case in which rapid cooling
dominates the dynamics of the gas. If perturbations are able to cool and
collapse more rapidly than they can sink, condensation proceeds on a time scale
$\sim \tcool$. When global thermal balance is maintained, the average
$\TimeScaleRatio$ of the ambient gas quickly increases as condensation lowers
the gas mass and density of the ambient medium. Within a few cooling times,
$\TimeScaleRatio$ rises to $\gtrsim 10$, as shown in Figure
\ref{fig:timescale_short}. In this strong-cooling regime, the onset of
condensation is determined by the growth of the initial perturbations and does
not depend strongly on the initial ratio of $\TimeScaleRatio$. Condensation
continues unabated until the timescale ratio is above 10, at which point the
cooling is weak enough that the condensation rate slows.\par

\subsection{Instability Growth in the Convective Regime}
\begin{figure}
   \includegraphics[width=0.5\textwidth]{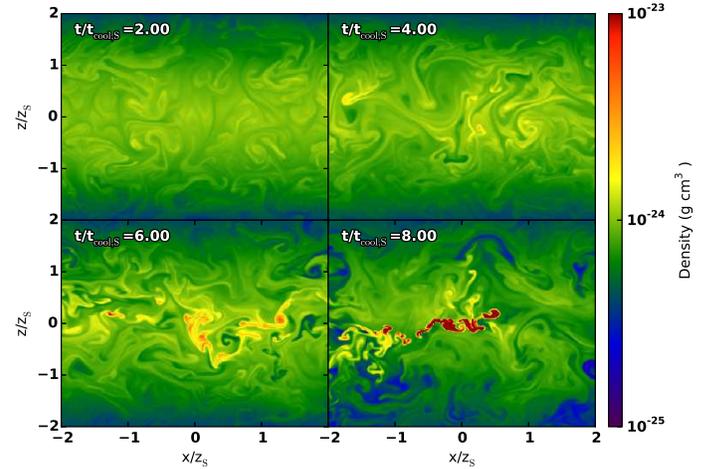}
   \centering
   \caption{Evolution of gas density in a 2D planar simulation with an
   iso-cooling initial condition of $\TimeScaleRatio=5.0$.  After two cooling
   times, the gas is clearly convecting. At four cooling times no cold gas has
   condensed, but the amplitude of the perturbations has increased. The
   perturbations have been further amplified after six cooling times, and the
   first condensate has formed. After eight cooling times, the densest gas near
   the center has entered into runaway cooling, leading to continuous
   condensation.}
   \label{fig:time_evolution}
\end{figure}

When the initial ratio of $\TimeScaleRatio$ is large, incipient condensing
regions sink into the gravitational potential faster than they can cool, leading
to a roiling, convective state. The convection is subsonic, and although the
pressure remains nearly constant at a given height, convection does not prevent
the temperature and density perturbations generated by cooling from growing.
Figure \ref{fig:time_evolution} illustrates the growth of perturbations in a
medium with an initial timescale ratio of $\TimeScaleRatio=5.0$. After 2 cooling
times (1 cooling time = 585 Myr), the gas is convecting. After 4 cooling times,
the gas continues to convect, but the density perturbations have increased.
After 6 cooling times, convection can no longer suppress condensation of  gas
near the midplane, and it cools catastrophically. After 8 cooling times, a
significant amount of the dense gas has condensed.\par

\begin{figure}
   \includegraphics[width=0.5\textwidth]{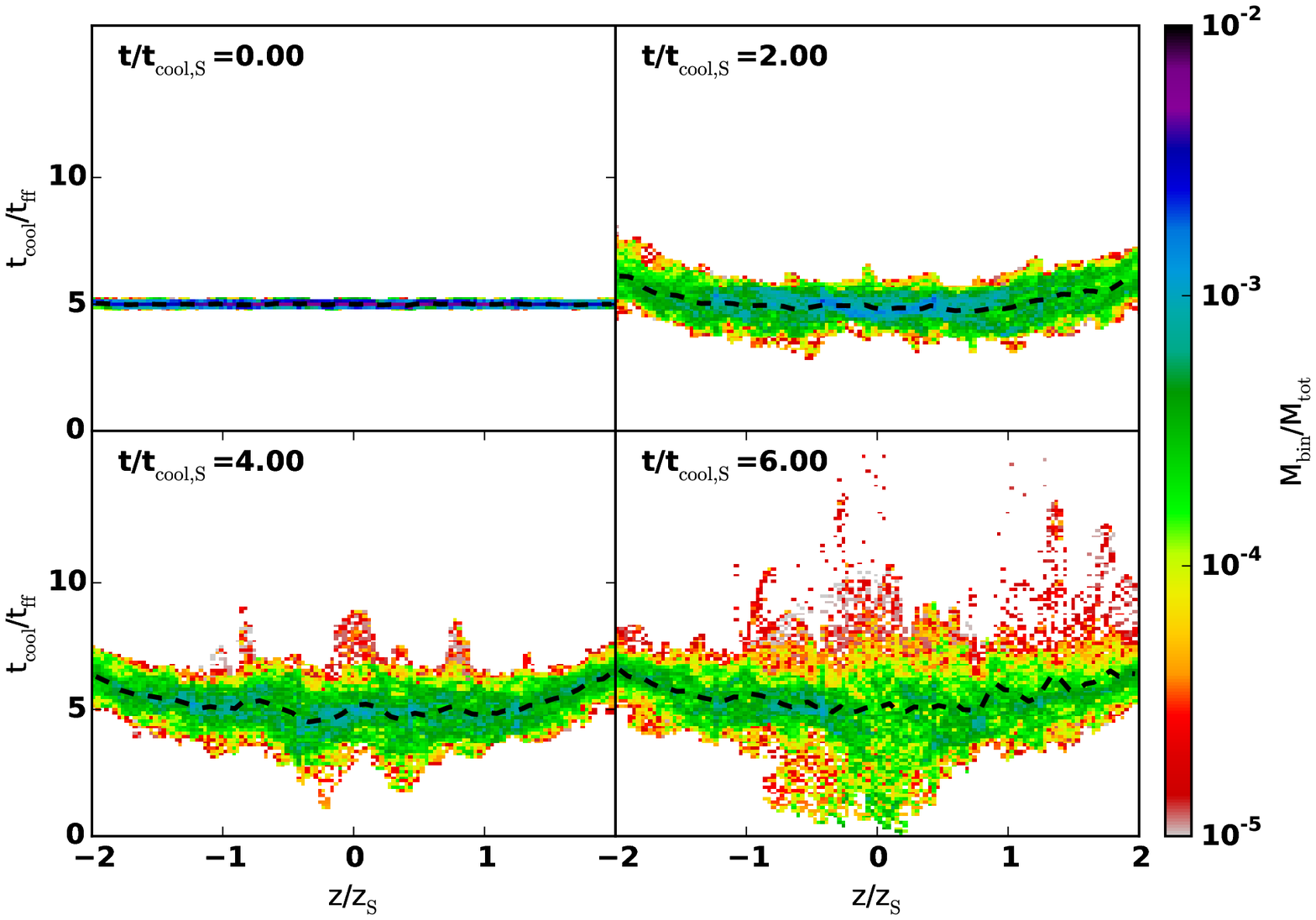}
   \centering
   \caption{Evolution of the mass-weighted probability distribution for the
   ratio of cooling time to freefall time for a 2D planar geometry with initial
   $\TimeScaleRatio=5.0$. The dashed black line shows the volume-weighted
   average ratio as a function of height. Note that when gas condenses, most of
   the volume is occupied by the hot gas, meaning that the volume-averaged ratio
   will tend to lie above the mass-weighted mean. The first panel shows the
   initial state of the gas where the timescale ratio is held constant
   throughout (with some spread due to the initial perturbation spectrum). At
   $t=2.0~t_{\rm cool,S}$, the gas has entered into a convective state and
   although condensation has not yet commenced, a spread in gas properties is
   evident. By $t=6.0~t_{\rm cool,S}$, a portion of the gas has reached a state
   with $\TimeScaleRatio \approx 2-3$, and the condensation process has begun.
   Although some gas is entering into the cold phase, the volume averaged ratio
   of $\TimeScaleRatio$ remains near its initial value as the cold gas occupies
   negligible volume.}
   \label{fig:phase_evolution_timescale_2}
\end{figure}

It is thus clear that the condensation does not simply switch on when the
average ratio of cooling time to the dynamical time drops below some special
value. To quantify the transition of the gas from a relatively smooth,
convective state to a multiphase medium, we plot in Figure
\ref{fig:phase_evolution_timescale_2} the probability distribution function of
the thermal state of the gas as the run with initial $\TimeScaleRatio = 5.0$
evolves. After several cooling times the distribution of gas in the
$z$--($\TimeScaleRatio$) plane has widened considerably. After 4 cooling times,
gas in the tail of the distribution has reached a ratio of around 3.  At this
point, further perturbation growth is inevitable and condensation begins.\par

\begin{figure}
   \includegraphics[width=0.5\textwidth]{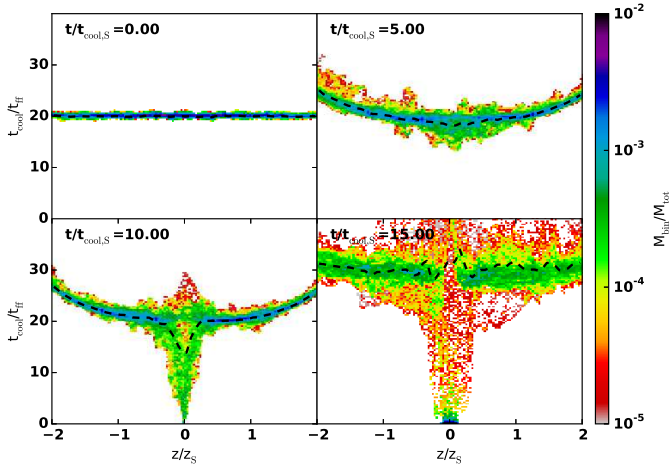}
   \centering
   \caption{Same as Figure \ref{fig:phase_evolution_timescale_2} except for an
   initial timescale ratio of $\TimeScaleRatio=20.0$. By $t=5.0
   \,\,\mathrm{t_{cool}}$, perturbations have started to grow but have not yet
   led to condensation. As the gas is able to undergo more freefall times per
   cooling time than in the case of $\TimeScaleRatio=5.0$, cooler gas is able to
   effectively settle towards the midplane. Nevertheless, condensation is still
   able to occur near the midplane even though the volume averaged value of
   $\TimeScaleRatio$ remains near 10.}
   \label{fig:phase_evolution_timescale_7}
\end{figure}

Increasing the initial timescale ratio to $\TimeScaleRatio = 20.0$ slows the
condensation process and further restricts it to the midplane region, as shown
in Figure \ref{fig:phase_evolution_timescale_7}. Condensation follows the same
general pattern as in Figure \ref{fig:phase_evolution_timescale_2}, except that
it is delayed for more than 10 times the initial cooling time and is much more
pronounced near the midplane. The concentration toward the midplane occurs
because cooling gas blobs can settle over a larger number of freefall times and
preferentially accumulate in the midplane before condensing.\par

\begin{figure}
   \includegraphics[width=0.5\textwidth]{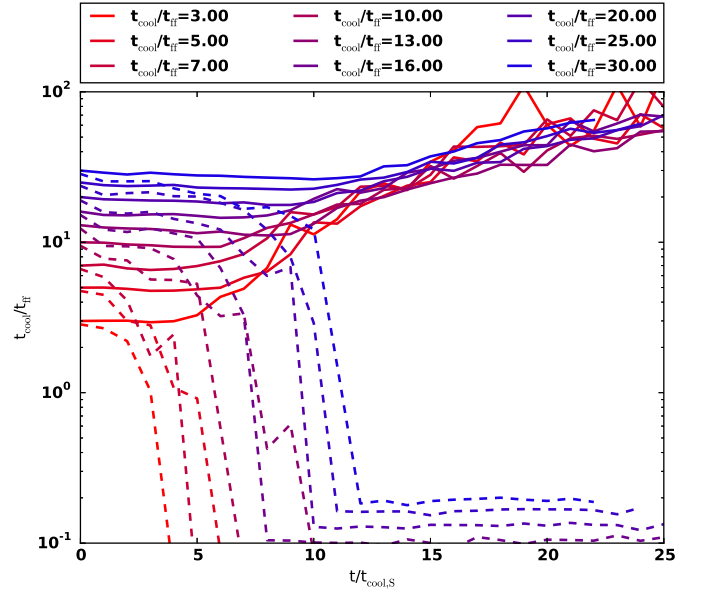}
   \centering
   \caption{Same as Figure \ref{fig:timescale_short}, except for runs with
   larger initial values of $\TimeScaleRatio$. In each simulation, the minimum
   value of $\TimeScaleRatio$ decreases on a timescale roughly proportional to
   the cooling time. When the initial ratio is higher, it takes several cooling
   times for gas to develop regions with a minimum timescale ratio near unity;
   therefore, condensation is delayed in these runs. Note that runs with larger
   values of $\TimeScaleRatio$ have a low overall cooling rate which, combined
   with the temperature floor of $\mathrm{T_{floor}}=5\times 10^6$ K, produces
   the floor in the timescale ratio.}
   \label{fig:timescale_vs_cooling_time}
\end{figure}

In all of our simulations, which have iso-cooling initial conditions up to
$\TimeScaleRatio = 30$, condensation eventually occurs as long as it is given
enough time to develop. Figure \ref{fig:timescale_vs_cooling_time} shows how
both the average and minimum values of $\TimeScaleRatio$ evolve during each run.
Condensation in the runs with large values of $\TimeScaleRatio$ may be
surprising in light of recent theoretical studies predicting that the medium
should become multiphase only if $\TimeScaleRatio \lesssim 10$
\citep{2012MNRAS.420.3174S,2012ApJ...746...94G,2015MNRAS.446.1895S}, and we will
discuss possible explanations for this difference in Section
\ref{section:discussion}.

\begin{figure}
   \includegraphics[width=0.5\textwidth]{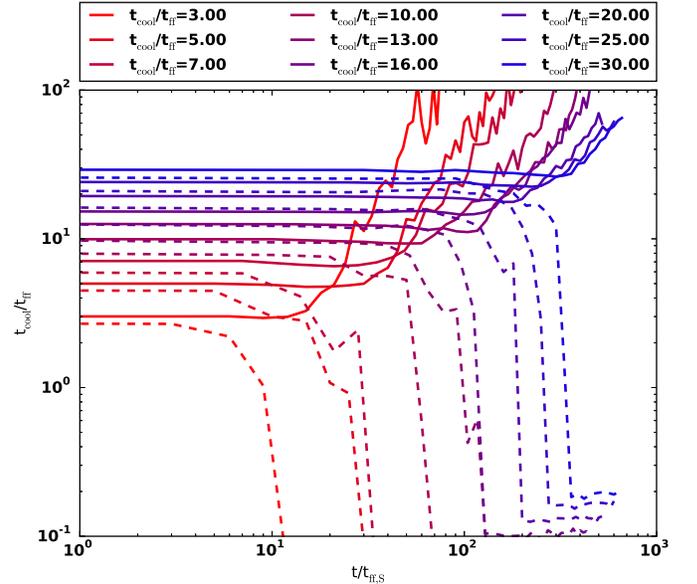}
   \centering
   \caption{Same as Figure \ref{fig:timescale_vs_cooling_time} except with time
   plotted in units of the freefall time at one scale height rather than the initial
   cooling time at one scale height. 
   }
   \label{fig:timescale_vs_freefall_time}
\end{figure}

Figure \ref{fig:timescale_vs_freefall_time} shows the same simulations as
Figure \ref{fig:timescale_vs_cooling_time} plotted with time in units of the 
freefall time at one scale height rather than the initial cooling time. 
As all simulations use the same gravitational potential, the freefall time
is a standard clock and corresponds to the same time interval in physical units (approximately
117 million years). When the initial ratio of cooling time to freefall time exceeds
$\sim 10$, condensation occurs after $\sim 100$ freefall times,
corresponding to a timescale comparable to the Hubble time. 

\subsection{Transition to the Condensed State}
\begin{figure}
   \includegraphics[width=0.5\textwidth]{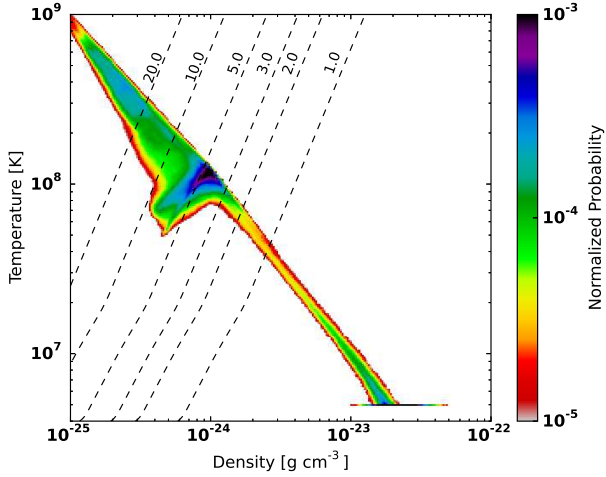}
   \centering
   \caption{The probability distribution function of the gas in the $\rho-T$
   plane, averaged over 30 cooling times in the 2D planar simulation with an
   initial timescale ratio of $\TimeScaleRatio=5$. Note that a large fraction of
   the gas is located at the temperature floor, near the x-axis at a density
   slightly above $10^{-23}$ g cm$^{-3}$. As the convection and condensation
   processes proceed subsonically, the process is largely isobaric, with a
   modest spread due to gravitational stratification. Lines of constant cooling
   time are shown as dashed lines, labeled with the ratio of cooling time to
   freefall time at 1 scale height. Note that the gas spends very little time
   between the line $\TimeScaleRatio=2$ and the temperature floor, indicating
   that once the threshold is reached, condensation proceeds rapidly.}
   \label{fig:time_weighted_plot}
\end{figure}

While studying the growth of thermal instabilities gives insight into the
conditions under which gas will condense, it does not necessarily explain how
the gas reaches a condensed state or how an individual parcel of gas behaves.
Figure \ref{fig:time_weighted_plot} depicts the gas distribution in the $\rho-T$
plane integrated over 30 cooling times. To compute this distribution, we bin gas
mass in $\rho-T$ space in each data output (which are evenly spaced in
time), sum over all of the
outputs, and normalize so that the integral over the distribution is equal to 1.
This probability distribution corresponds to the probability of a parcel of gas
being found in a given thermodynamic state at some point during the simulation.
The gas is for the most part constrained to a line of constant pressure with
spread due to gravitational stratification. The distribution has two peaks; a
low density, high temperature node in which the gas is convecting, and a cool,
low temperature node representing the condensed state of the gas. The
probability of finding gas in the connecting region is low, indicating that
condensation from the hot phase into the cold phase proceeds rapidly once it
begins.\par

\begin{figure}
   \includegraphics[width=0.5\textwidth]{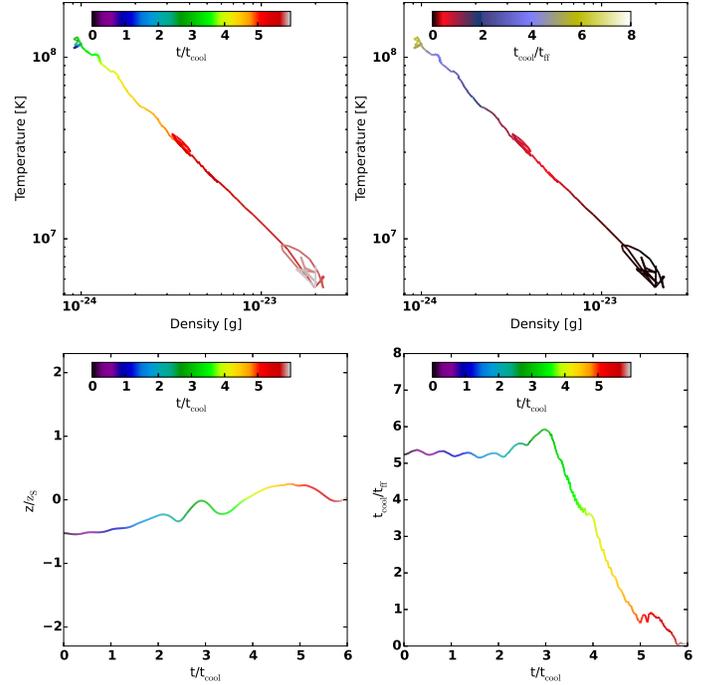}
   \centering
   \caption{The dynamics of fluid during the condensation process are shown in
   the dynamics of a Lagrangian tracer particle through phase space. The
   particles are inserted during initialization in the 2D planar simulation with
   an initial timescale ratio of 5. The upper left panel shows the particle's
   path through $\rho-T$ space, with the color of the line showing elapsed time
   in cooling times. The upper right panel also shows the path through $\rho-T$
   space, but is colored by the ratio of cooling to freefall time. The bottom
   left panel shows particle height vs. time, and the bottom left shows the
   timescale ratio as a function of time.}
   \label{fig:tracer_comparison}
\end{figure}

Figure \ref{fig:tracer_comparison} illustrates the dynamics of the gas during
the convective stage and the condensation process using the motion of a
Lagrangian tracer particle in phase space. The figure shows the path of a
representative particle which condenses early in the simulation. For several
cooling times, the gas simply convects within a narrow portion of phase space.
As the thermal perturbations are amplified, the gas is driven to a colder,
denser state which is where condensation occurs. When the gas does condense, the
condensation process is very rapid, and the gas stays in the condensed phase
afterwards.\par

\subsection{Condensation Rate}
\begin{figure}
   \includegraphics[width=0.5\textwidth]{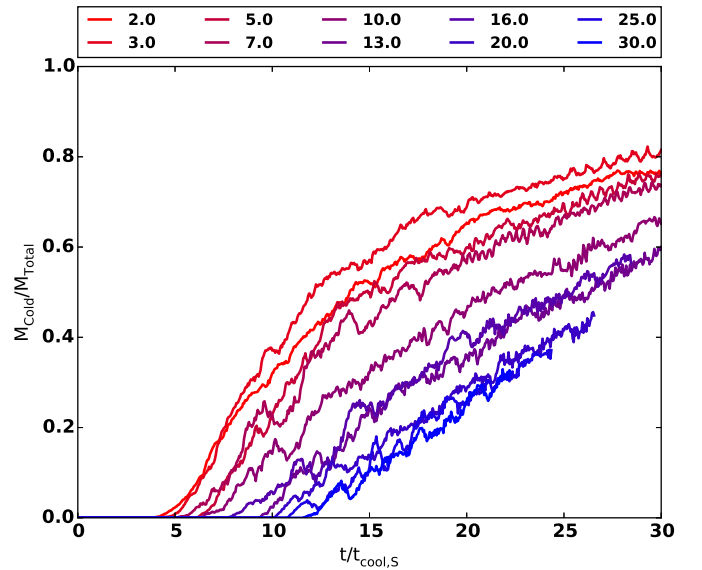}
   \centering
   \caption{The fraction of mass in the condensed state is shown as a function
   of time for 2D planar runs with large initial values of $\TimeScaleRatio$.
   The condensed fraction is measured over the entire domain.}
   \label{fig:cold_mass_vs_t_cool}
\end{figure}

In our simulations, condensed gas remains in the condensed state and settles
towards the center. After the onset of condensation the gas segregates into two
phases - the cool condensed material in the center and the hot, convective gas
that remains uncondensed. This departure from the expectation of self-regulation
is a consequence of our feedback implementation and is discussed further in
Section \ref{section:self_regulation}. Still, it is instructive to examine the
rate of condensation in our simulations, as is shown in Figure
\ref{fig:cold_mass_vs_t_cool}. Following the onset of condensation,
instabilities continue to grow on the cooling timescale. Each simulation behaves
similarly on a thermal timescale, with a roughly linear growth in the total
condensed fraction.\par

\subsection{Effect of Geometry}
\begin{figure}
   \includegraphics[width=0.5\textwidth]{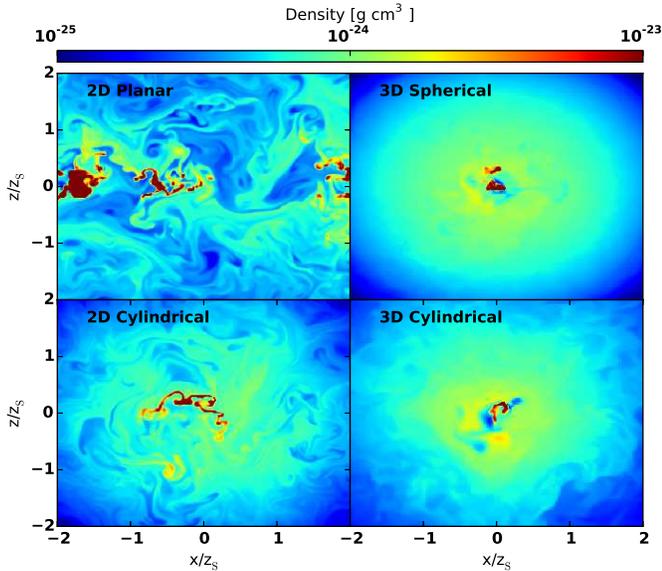}
   \centering
   \caption{The evolution of runs with an initial $\TimeScaleRatio$ of 5.0 are
   shown for different geometries at $t=10\,\tcool$. All runs use an identical
   setup with respect to the radial coordinate, though the definition or the
   radial coordinate is changed based on the geometry of the simulation. The 2D
   runs use a static grid of $300 \times 300$ cells, while the 3D runs use a
   root grid of $128^3$ cells with 2 layers of adaptive mesh. The slices of
   density through the cylindrical runs are taken perpendicular to the axis of
   symmetry.}
   \label{fig:geometry_comparison}
\end{figure}

Simulations with different geometries are shown in Figure
\ref{fig:geometry_comparison}. All simulations use the same initial conditions
($\TimeScaleRatio = 5.0$). In the spherical case, gravity pulls towards the
origin, while in the cylindrical setups gravity pulls towards the symmetry axis.
All simulations exhibit similar thermal behavior. After 10 cooling times, the
gas has entered into a convective state and condensation has begun near the
center of the potential. In the non-planar runs, less gas condenses as the
region near the center occupies less volume. Nevertheless, we do not observe a
significant change in the condensation process among simulations with different
geometries.\par

%% file: discussion.tex
\section{Discussion and Relationship to Related Work}\label{section:discussion}
Our simulations would seem to indicate that any medium subject to a
heating/cooling balance as we have described in our model will eventually
succumb to thermal instability and produce condensation. Nevertheless,
observations seem to indicate that clusters with time scale ratios above roughly
10 do not produce much multiphase gas. To explain this discrepancy, we note that
at a radius of around 30 kpc, a time scale ratio of 10 in a large galaxy cluster
corresponds to a cooling time on the order of a Gyr. Physical processes such as
mergers, star formation, and AGN feedback occur on shorter timescales, rendering
the condensation process sub-dominant in these cases.  Therefore, in a realistic
cluster environment only clusters with a cooling time to freefall time ratio of
$\lesssim 10$ are likely to develop condensation. An important caveat to this
observation is that while the growth of thermal instabilities from initially
small perturbations may be unimportant on cluster timescales, if the gas is
inhomogeneous due to other physical processes (such as an AGN jet) condensation
may occur in the tail of the thermal distributions shown in Figure
\ref{fig:phase_evolution_timescale_2} and \ref{fig:phase_evolution_timescale_7}.
Thus, predicting the onset of condensation is not as simple as measuring the
value of $\TimeScaleRatio$; the level of inhomogeneity must also be taken into
account.\par

Our simulations examine the formation of multiphase gas in an idealized setting
wherein global balance between heating and cooling is strictly enforced.  While
this model gives rise to results that are qualitatively consistent with
observations, it clearly neglects the complex physics of AGN feedback and heat
transport which occur in real clusters. In this section, we discuss our results
in light of current observations of multiphase gas and previous simulations of
condensation and consider the complications that inclusion of additional
physical processes would cause.\par

\subsection{Observations of Multiphase Gas}
Owing to the timescales involved and the limits of current telescopes,
astronomers can not directly observe the condensation process in the ICM.
Nevertheless, the past decade has deepened the field's appreciation of a
fascinating dichotomy in cluster properties when cluster cores are probed for
cold gas and signatures of AGN feedback. While cooling is generally suppressed
in cool-core galaxy clusters \citep{2003ApJ...590..207P, 2006PhR...427....1P},
at least some cold gas is observed in galaxies with low central temperatures, as
seen in the works of \citet{2010ApJ...721.1262M} and
\citet{2010MNRAS.407.2063W}.  \citet{2008ApJ...683L.107C} considers the entropy
profiles, radio emissions, and presence of H$\alpha$ in the ACCEPT sample of 222
galaxy clusters.  As H$\alpha$ emission requires the presence of cold (relative
to the ICM) gas, the presence or absence of H$\alpha$ in a cluster may be taken
as an indicator of multiphase gas. In the clusters with H$\alpha$ observations,
H$\alpha$ is conclusively detected in slightly over half of the sample. A
strong correlation is seen between the presence of H$\alpha$ and the core
entropy; clusters with H$\alpha$ have central entropies below 30 keV cm$^2$,
while those without H$\alpha$ detections tend to lie above the 30 KeV line. When
the entropy profile is used to infer a cooling time, \citep[as
in][]{2015ApJ...799L...1V,2014arXiv1409.1598V} a central entropy of 30 KeV
corresponds to a cooling time of around 1 billion years, consistent with a
cooling time to freefall time ratio of around 10. In clusters in the ACCEPT
sample, those with detected H$\alpha$ emission consistently have
$\TimeScaleRatio$ values below $\simeq 20$, while those without H$\alpha$
detections lie entirely above that value.\par

\subsection{Simulations of Multiphase Gas}
\citet{2012MNRAS.419.3319M}, upon which this study is based, finds that
precipitation will occur rapidly if the gas is able to cool in place, which
occurs when $\TimeScaleRatio \lesssim 1$. The authors also conclude that the
condensation process is relatively insensitive to variations in the heating rate
and mechanism. Employing a similar method, \citet{2012MNRAS.420.3174S} finds
that condensation may occur in gas with a timescale ratio of $\lesssim 10$ in a
spherical simulation, an enhancement they attribute to the compression of
overdense blobs descending in a spherical geometry. Additionally,
\citet{2012MNRAS.420.3174S} concludes that condensation does not occur when the
timescale ratio rises above 10.\par

Analytic work has lent further credence to the idea that $\TimeScaleRatio
\lesssim 10$ represents a critical threshold for condensation.
\citet{2015MNRAS.446.1895S}, extending the analysis of
\citet{2005ApJ...632..821P}, finds that small instabilities may grow when
$\TimeScaleRatio \lesssim 1$ for planar geometries and, when the effects of
geometric compression are included, may grow for $\TimeScaleRatio \lesssim 10$
for spherical geometries. While the results presented in Section
\ref{section:results} of this paper suggest a moderately higher threshold for
the planar case, we believe that these results are largely consistent with
\citet{2015MNRAS.446.1895S} in the context of individual overdensities cooling
and condensing in place. In our simulations, however, we see that overdensities
in a medium above the critical threshold oscillate, leading to a roiling state
that develops further perturbations. This cross talk effect between layers
generates non-linear perturbations and causes the temperature dispersion in the
medium to grow on the cooling timescale.  When the cold tail of the distribution
has dropped to $\TimeScaleRatio \lesssim 2-3$ the condensation process begins.
Thus we find that the mechanism responsible for condensation above the critical
threshold is not geometric compression but the continued growth of perturbations
following the onset of convection in the gas.\par

Simulations that employ more realistic heating mechanisms also find that
condensation occurs in galaxy clusters, albeit under somewhat
different circumstances than
in simulations with idealized heating. \citet{2014ApJ...789..153L} employ an
AGN feedback algorithm in which heating is triggered by cold gas accretion. The
study finds that condensation occurs when $3 \lesssim \TimeScaleRatio \lesssim
10$. This condensation occurs along the axis of the jet, where dense gas is
dragged up and is able to cool as it falls. This is consistent with our
findings, in which the thermal instability can grow when $\TimeScaleRatio
\lesssim 10$, but only when gas is sufficiently hetrogeneous.  Similarly,
\citet{2012ApJ...746...94G} employs jet heating in response to accretion and
finds that multiphase gas can form when $\TimeScaleRatio \lesssim 10$.\par

\subsection{Caveats and Limitations}\label{section:caveats}
In this study, we have used an idealized model to simulate the onset of
condensation in galaxy clusters. While the simplicity makes this model easy to
analyze, we have left out physics that may have significant impact on the
development of a multiphase medium. In particular, conduction and the presence
of magnetic fields may inhibit or shape the growth of condensation. Conduction
works to smooth out temperature perturbations, while magnetic fields will lead
to conduction being anisotropic.\par

Magnetic fields in clusters are poorly understood. While weak, they are known to
be present and may be dynamically important in cluster cores
(\citet{2002ARA&A..40..319C} and references therein). More important for this
work, magnetic fields in a plasma will lead to anisotropic conduction,
channeling heat along the direction of magnetic field lines as explored in
\citet{2011ApJ...740...81R}. Similarly, \citet{2014MNRAS.439.2822W} studies the
growth of thermal instabilities in a spherical setup and includes both
conduction and magnetic fields. Anisotropic conduction is not found to inhibit
condensation, but does lead to the formation of filaments rather than globules
of dense gas.  Conduction is found to inhibit condensation if the efficiency is
above 0.3 of the Spitzer value.\par

In addition to these omissions, our assumed heating function does not capture
the true physical process responsible for transferring energy from the AGN to
the ICM. While the details of AGN heat transfer are not currently understood,
several mechanisms have been proposed, including shocks
\citep{2012NJPh...14e5023M,2004ApJ...611..158R}, cosmic
rays \citep{2010ApJ...720..652S,2013MNRAS.432.1434F,2011ApJ...738..182F,2012ApJ...746...53F},
turbulent mixing \citep{2009ApJ...699..348S,2014MNRAS.443..687B}, PdV work from
the inflation of hot bubbles \citep{2007ARA&A..45..117M,2004ApJ...607..800B},
and the uplifting of cool gas by rising bubbles \citep{2010MNRAS.407.2046M}. The
actual heating function is unlikely to maintain perfect thermal balance, and
presumably does not act in a strictly volumetric sense as assumed in this work.
Still, the lack of cold gas and star formation in cool-core clusters implies
that the heating function must broadly maintain thermal equilibrium, making the
model considered in this work physically relevant.

\subsection{Self-Regulation} \label{section:self_regulation}
The gas does not reach a steady state, as might be expected for an ideal
self-regulating system. Instead, condensation continues in the convective gas
after the condensation process has begun, increasing the separation between the
hot and cold phases. In real clusters, feedback is expected to operate in a
thermostat-like manner, which should produce a rough thermal equilibrium. The
lack of self-regulation in our simulations is purely an effect of the heating
model that we employ, and does not accurately capture the response of feedback
to condensation. However, if we imagine feedback to be powered by condensation,
we can use the calculated heating rate to determine what feedback efficiency
would be necessary for the system to balance radiative losses.\par

\begin{figure}
   \includegraphics[width=0.5\textwidth]{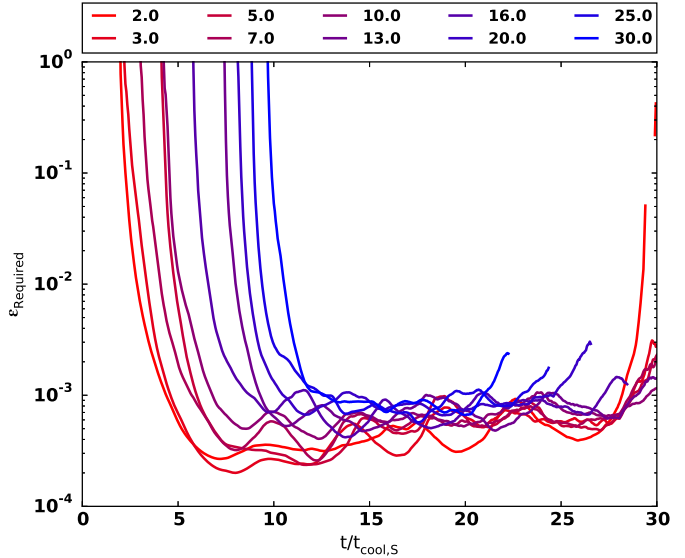}
   \centering
   \caption{The feedback efficiency necessary to maintain thermal balance in the
   hot gas is shown for 2D planar runs starting with different values of
   $\TimeScaleRatio$. The required efficiency is calculated as
   $\epsilon=\dot{E}/ \dot{M}_{Cold}c^2$, where $\dot{E}$ is the cooling rate of
   all gas above the temperature floor. Both the cooling rate and the
   condensation rate have been smoothed over a cooling time.}
   \label{fig:feedback_efficiency}
\end{figure}

During accretion, AGN are expected to convert a significant fraction of the
infalling mass into energy that is then returned to the surrounding medium. The
feedback rate can be related to the mass accretion via

\begin{equation}
   \dot{E} \approx \epsilon \dot{M} c^2
\end{equation}

\noindent where $\dot{E}$ is the total energy output, $\epsilon$ is an efficiency
parameter, and $\dot{M}$ is the mass accretion rate. Under the assumptions that
all of the condensing gas is used to power feedback and that all of the feedback
energy is transferred to the ICM, we have estimated the conversion efficiency
necessary to maintain thermal balance. The estimate is shown in Figure
\ref{fig:feedback_efficiency} for several initial values of $\TimeScaleRatio$.
Once condensation has begun, the required efficiency in all runs reaches a value
of around $10^{-3}$, in line with the values found in
\citet{2012MNRAS.420.3174S}. As the accumulation of cold gas near the midplane
is an artifact of our setup and would not be expected in a real cluster, we
calculate the cooling rate over the ambient gas, which is that gas that is above
the temperature floor of $5.0 \times 10^6$ K. Although we do not explore the mechanism for releasing
mass-energy from condensed gas in this work, if condensation resulting from the
growth of thermal perturbations is in principle capable of balancing radiative
cooling in the ICM, thermal instability must be taken seriously as a feature of
a self-regulating energy cycle in cool-core clusters.\par

%% file: conclusions.tex
\section{Conclusions and Future Work}\label{section:conclusions}
In this study, we have investigated the onset of convection in a thermally
unstable medium using an idealized model, including a heating scheme that  
strictly enforces a global heating-cooling balance. Although a simplification,
this model gives insight into the conditions necessary for the onset of
condensation in a gravitationally stratified medium such as that in a cool-core
galaxy cluster. This study indicates that condensation proceeds as follows:

\begin{itemize}
   
   \item If heating is able to balance cooling at all radii, thermal
instabilities will grow in amplitude, regardless of the initial
conditions.
      
   \item If the ratio of the cooling to the freefall time is $\lesssim
2$, (the strong cooling regime) the gas will condense in place,
driving the volume-averaged $\TimeScaleRatio$ value above 10.
   
   \item Above a ratio of $\TimeScaleRatio \approx 10$, perturbations
will grow on a timescale proportional to the cooling time.
   
   \item Once the perturbation distribution has broadened, gas with
$\TimeScaleRatio \approx 2-4$ will condense, even if the
volume-averaged ratio of $\TimeScaleRatio$ is above 10.
   
   \item If the timescale ratio is $\gtrsim 10$, the timescale for
condensation to occur in gas with $\tcool \sim 1$~Gyr is comparable to
the Hubble Time and greatly exceeds other relevant cluster timescales.
   
\end{itemize}

A fundamental limitation of this work is that the model assumes a heating
function that is idealized and does not mimic a specific physical process. In
preparation for future work, it will be necessary to examine a greater variety
of heating modes, including models more analogous to jet feedback and quasar
winds from accreting supermassive black holes. The physical processes underlying
black hole accretion, feedback, and heat transfer to the ICM are still poorly
understood, and elucidating them will form the focus of future studies.\par

%% file: acknowledgments.tex
\acknowledgments
\section{Acknowledgments}
The authors would like to thank Gus Evrard, Devin Silvia, Greg Bryan, and Yuan
Li for helpful discussions during the preparation of this paper. This work was
supported by NASA through grant NNX12AC98G and Hubble Theory Grant
HST-AR-13261.01-A, and by the National Science Foundation through grant
PHY-0941373. The simulations presented in this paper were performed and analyzed
on the TACC Stampede supercomputer under XSEDE allocations TG-AST090040 and
TG-AST100004. This work was supported in part by Michigan State University
through computational resources provided by the Institute for Cyber-Enabled
Research. BWO was supported in part by the sabbatical visitor program at the
Michigan Institute for Research in Astrophysics (MIRA) at the University of
Michigan in Ann Arbor, and gratefully acknowledges their hospitality.
\texttt{Enzo} and \texttt{yt} are developed by a large number of independent
researchers from numerous institutions around the world. Their commitment to
open science has helped make this work possible.